\documentstyle[aps,prb,preprint,tighten]{revtex}
\begin{document}
\draft
\title{Phenomenological Models for the Gap Anisotropy of Bi-2212\\
as Measured by ARPES}
\author{M. R. Norman and M. Randeria}
\address{Materials Science Division,
Argonne National Laboratory, Argonne, Illinois  60439}
\author{H. Ding and J. C. Campuzano}
\address{Materials Science Division,
Argonne National Laboratory, Argonne, Illinois  60439
and Department of Physics, University of Illinois at Chicago,
Chicago, Illinois  60607}
\maketitle

\begin{abstract}

Recently, high resolution angle-resolved photoemission spectroscopy has
been used to determine the detailed momentum dependence of the
superconducting gap in the high temperature superconductor Bi-2212.
In this paper, we first describe tight binding fits to the normal state
dispersion and superlattice modulation effects. We then discuss
various theoretical models in light of the gap
measurements.  We find that the simplest model
which fits the data is the anisotropic s-wave gap
$\cos(k_x)\cos(k_y)$, which within a one-band
BCS framework suggests the importance of  next near neighbor Cu-Cu
interactions. Various alternative interpretations of the observed
gap are also discussed, along with the implications
for microscopic theories of high temperature superconductors.

\end{abstract}

\bigskip


\narrowtext

\noindent{\bf 1. Introduction}

One of the key issues today in the field of high temperature superconductivity
is the symmetry of the superconducting order parameter.  Penetration depth
measurements\cite{hardy} on YBCO indicate that the gap has line nodes in
momentum space.  This information by itself is not sufficient to determine
the symmetry of the order parameter.  Various Josephson
interference experiments \cite{vh,ring} on YBCO are consistent
with an order parameter which has $d_{x^2-y^2}$ symmetry;
some of these directly probe the sign change in the
the order parameter under a 90 degree rotation.\cite{vh}
These results are, however, not consistent
with c-axis Josephson tunneling data\cite{dynes} and other interference
experiments.\cite{chaud}  The situation is made even more confusing by
experiments on other high $T_c$ materials:
penetration depth measurements\cite{anlage} and tunneling data\cite{jz}
on NCCO and tunneling data on the single layer Hg cuprate\cite{jz2}
indicate an apparently isotropic gap.

The improvement in resolution of angle-resolved	photoemission spectroscopy
(ARPES) as well as the large gaps associated with the high transition
temperatures of the cuprates allows for the first time the possibility of
directly mapping out the momentum dependence of the gap.  We are indeed
fortunate that the (quasi) two-dimensional nature of the cuprates
allows ARPES to directly measure the spectral function of the electrons.
Recently, the Stanford group \cite{shen}
presented the first ARPES data indicating significant momentum anisotropy
in the gap function for the two layer Bi cuprate (Bi-2212),
with the gap being large along the CuO bond direction $(\pi,0)$ and
small, possibly zero, in the diagonal direction $(\pi,\pi)$ in contrast to
early ARPES data\cite{olson} on Bi-2212 which
showed no gap anisotropy.  More recent work \cite{ding,onel} has confirmed
the Stanford results and found that
the observed gap anisotropy is sensitive to sample quality
and surface conditions \cite{ding} with some samples showing
a much smaller, though non-zero,\cite{onel} gap along
the $(\pi,\pi)$ direction compared with that along $(\pi,0)$.
Even in the Stanford data the anisotropy
decreased significantly with sample aging, probably
due to gas absorption on the surface.\cite{shen}
It must be mentioned that
attempts to see the gap with ARPES on other cuprates, particularly YBCO,
have not been successful.  The reason appears to be that in Bi-2212, the
cleavage is between the two BiO layers which are van der Waals coupled.
Thus, the act of cleavage minimally disrupts the sample surface, as opposed
to YBCO where chain copper $-$ apical oxygen bonds are broken.\cite{bansil}

Recently, our group has measured the detailed momentum
dependence of the gap in Bi-2212 on very high quality samples\cite{sample}
employing a spectrometer with improved energy
resolution (described by a gaussian of standard deviation 8 meV).\cite{gap}
These results are consistent with the earlier work
discussed above, but with improved resolution, were able to demonstrate that
the nodes in the gap function were not along the $(\pi,\pi)$ directions as one
would have for a gap of $d_{x^2-y^2}$ symmetry, but rather displaced at an
angle of 10 degrees to both sides of the diagonal direction.  The purpose of
this paper is to discuss the detailed angle dependence of the gap, and see
which theoretical models are consistent with such a gap function.

Our main results are summarized below. \hfill\break
(1) We give a tight binding fit to the normal state dispersion data
which reproduces the experimentally observed Fermi surface.\hfill\break
(2) We note that there is no clear evidence for two CuO bands, suggesting
that the bilayer splitting is either weak or non-existent.\hfill\break
(3) The occupied area of the Fermi surface corresponds to a hole
doping of 17\%. \hfill\break
(4) We demonstrate that data in the Y quadrant are consistent with
the superlattice modulation observed in structural studies. \hfill\break
(5) The observed momentum dependence of the superconducting gap is not
compatible with pure $d_{x^2-y^2}$ pairing. We also argue that the
data are not consistent with a dirty $d$-wave or a mixed ($s+d$, $s+id$,
or $d+g$) gap. \hfill\break
(6) The observed gap in the X-quadrant
is consistent with an interplay between a pure $d_{x^2-y^2}$ state
and the superlattice modulation. However this interpretation
is not consistent with the data in the Y-quadrant. \hfill\break
(7) A mixed $s \pm d$ gap coming from pairing within a bilayer
can fit the data but at the expense of having two gaps at each k point.
So far there is no evidence in our data for such a two-gap
spectrum.\hfill\break
(8) The simplest interpretation of the data is in terms of
anisotropic s-wave pairing. An $s_{xy}$
gap function with $\Delta({\bf k}) = \Delta_0\cos(k_x)\cos(k_y)$
provides an excellent fit to the data.\hfill\break
(9) $s_{xy}$ pairing would arise within a BCS framework from
next-near-neighbor (Cu-Cu) attraction with weak on-site repulsion.\hfill\break
(10) Several microscopic theories which lead to anisotropic
s-wave pairing are discussed in relation to the data,
including models based on charge transfer,
extended saddle points, and interlayer tunneling.

\noindent{\bf 2. Normal State}

We first discuss the normal state data. We will use the notation
$\bar{M} = (\pi,0)$ and $Y = (\pi,\pi)$ with $\Gamma = (0,0)$
where $\Gamma-\bar{M}$ is along the CuO bond direction.
The circles and crosses in
Fig. 1 are the Fermi surface crossings in the ARPES measurements in the normal
and superconducting states, respectively, and
the thick line is the Fermi surface of the tight binding fit
described below.  We used six tight
binding functions to fit the normal state dispersion data (the latter
determined by the peak positions of the ARPES spectra),
with the functions and their coefficients
listed in Table 1.  The six conditions used to determine the coefficients
were (1) two
points from the measured energy dispersion in the $\Gamma Y$ direction,
(2) two points from the measured dispersion in the $\Gamma \bar{M}$ direction,
(3) the measured Fermi surface crossing in the $\bar{M} Y$ direction,
and (4) the energy at the $Y$ point.  The latter is not experimentally
determined (since it corresponds to an unoccupied state)
but is a necessary constraint to obtain a good fit.
The criteria used was to note that the energy dispersion below
the Fermi energy looks like that of band theory,\cite{band}
just reduced by a factor of two, as previously noted by
Olson's group.\cite{olson2}  Therefore, we assumed
the same (i.e., half the band theory value)
to be true above the Fermi energy to obtain the energy at
the Y point.  The resulting energy dispersion fit is shown in Fig. 2.  This
looks similar to that inferred by the Stanford group\cite{shen2} with
two differences.  First, the energy of the $\bar{M}$ point is definitely below
the Fermi energy ($-$34 meV), and, second, only one CuO band is seen, not
the two that might have been expected due to the two CuO planes/unit cell.
These differences are primarily due to the increased energy resolution
of the present experiments.

This indicates that
the bilayer splitting is either very weak or non-existent.  Although this
is consistent with band theory results at the Fermi energy,
band theory in addition predicts a sizable
splitting at the $\bar{M}$ point related to interactions with BiO bands.
Such a splitting and the existence of BiO bands below the Fermi energy does
not appear to be consistent with our data.  Absence of bilayer splitting,
though, is consistent with theories based on incoherent c-axis
transport.\cite{pwa}
We contrast this with YBCO, where significant bilayer
splitting has been seen in ARPES measurements\cite{kaz} (consistent
with band theory).  This difference is due to the presence of chains in YBCO
which couple differently to the even and odd combinations of the bilayer,
which together with the buckling of the CuO layers, causes a sizable
splitting.\cite{oka}

We also note from Fig. 1 the presence of side sheets in the $Y$ quadrant.
The Q vector connecting the side sheets to the main sheet is within our
resolution equal to the superlattice Q vector, $0.21(\pi,\pi)$, seen in
structural experiments.\cite{with}  The lines going through the
side sheets are simply displacements of the tight binding fit by this Q vector.
We find that all of the data is consistent with this picture, and it is not
necessary, at least near the Fermi energy, to invoke the 2 by 2 modulation
proposed by Aebi et al.\cite{aebi} (we note that the samples of that group
come from the same source as our own).  In principle, more side sheets than
two will occur, but presumably the intensity of these higher order umklapps are
reduced by matrix element effects.

Finally, we note that the area
occupied by the observed Fermi surface is equivalent to a hole doping level of
17\%, the same as that for optimal $T_c$ in LSCO.  Using quoted values
for the stoichiometries of the cations for our samples \cite{sample}
($Bi_{2.17}Sr_{1.77}Ca_{1.01}Cu_{2.05}O_{8+x}$) and the above hole count,
we estimate an $x = 0.08$, which is consistent with a variety
of published results.

\noindent{\bf 3. Superconducting State}

The method for extracting the gap has been discussed in detail in our
earlier paper.\cite{gap}
The Fermi momenta are obtained by finding the minimum separation of the
quasiparticle peak from the chemical potential (operationally,
when the leading edge of the energy distribution curve has the maximum slope).
Experimentally the peak first approaches the
chemical potential, then disperses away, with the intensity decreasing
rapidly as expected in BCS theory.

The spectral function is assumed to be that of
BCS theory with a phenomenological linewidth to broaden the delta functions.
Fortunately, this linewidth does not enter the fits at sufficiently
low temperatures since the
experimental spectra are resolution limited
at 13 K where the gaps are determined.
The spectral function is multiplied by the Fermi occupation factor
and then integrated over momenta using the $\pm 1$ degree width of
the analyzer window and convolved with the observed energy resolution function.
The momentum integrations utilize the dispersion derived from the
tight binding fit.
The background contribution affects the fits shown in Ref.\ \onlinecite{gap}
only at binding energies larger than 80 meV and thus is irrelevant
to the gap determination.

Besides the overall intensity, representing unknown matrix elements
and normalization, this leaves only one adjustable parameter
in the fit: the absolute value (magnitude) of the gap.
The resulting gap values are plotted in Fig.~3.  The basic point to note
is that the data are consistent with nodes in the order parameter
about 10 degrees away from, and on both sides of, the $(\pi,\pi)$ directions.

The data differ somewhat in the two quadrants,
although it should be noted that the data in the
two quadrants were taken on different samples.  This is because the analyzer
does not have enough angular range to cover both quadrants in the geometry
used.  Nonetheless, repeated measurements on different samples yield consistent
results for the momentum dependence of the gap, with the location of the nodes
in the two quadrants being equivalent.  We also note that in any particular
quadrant, the measured gap has some sample dependence, as found in all ARPES
work.

\noindent{\bf 4. $d$-Wave and Related Models}

In this Section we discuss the extent to which our results
are compatible or incompatible with $d$-wave pairing.
Some of the experimental evidence for and against $d$-wave
in YBCO was described in the Introduction.
{}From a theoretical point of view it has been established \cite{scalapino}
that if antiferromagnetic spin fluctuations mediate pairing
then it must have $d_{x^2-y^2}$ symmetry.
Experimentally the spin fluctuations in Bi-2212 have
not been studied as intensively as in YBCO, nor are there
any phase coherence experiments in Bi-2212.
Recent NMR experiments \cite{takigawa} show a rapidly
decreasing Knight shift below $T_c$, giving strong evidence for
singlet pairing. The low temperature Knight shift and relaxation
data suggest a gapless state which could arise from
$d$-wave with impurity scattering, but gapless $s$-wave due to
magnetic impurities cannot be ruled out.

Returning to the results of Fig.~3 we see that
the data are inconsistent with a simple $d_{x^2-y^2}$ gap,
since by definition such a gap must vanish along
the $(\pi,\pi)$ directions where $k_x = k_y$.
Of course, Bi-2212 has a small orthorhombic distortion, which, even at $T_c$,
can lead to
mixing of $d_{x^2-y^2}$ with other symmetries.  In YBCO, $d_{x^2-y^2}$ can mix
with s-wave, but in Bi-2212, it can mix only with g-wave ($A_{2g}$
representation of the form $(x^2-y^2)xy$), and thus would still have nodes
along the diagonal
directions.  The difference occurs since in YBCO, the orthorhombic
axes are along the CuO bond directions, whereas in Bi-2212, they are along
the diagonal directions (and thus, reflection symmetry about the diagonal
directions is preserved unlike in YBCO).  Of course, $s-d$ mixing can still
occur in the non-linear gap equations below $T_c$, but the above symmetry
in Bi-2212 in a Ginzburg-Landau approach would imply two phase transitions
instead of one.\cite{kwj}
We also note that an $s+d$ gap simply moves the
node to one side or the other of the $(\pi,\pi)$ direction;
it cannot lead to nodes on both sides of $(\pi,\pi)$, whereas an $s+id$ gap
is nodeless.

Another alternative to consider is the dirty $d$-wave state
which would produce a gapless region around $(\pi,\pi)$.
Even though we do not feel that the error bars on the gap estimation
allow for this possibility, let us for the sake of argument
assume that the small gaps in a $\pm 10$ degree region about
the diagonal are all consistent with zero. Within the dirty $d$-wave
theory, when such a large region of gaplessness is produced around $(\pi,\pi)$
the large gap (near the $\bar{M}$ point here) would also be
suppressed. Given that this gap is experimentally found to be quite
sizable is further evidence against such an interpretation, although we
caution that the degree of suppression is dependent on whether one assumes
Born or unitary scattering.  A more definitive test of dirty d-wave would be to
conduct ARPES experiments as a function of disorder and compare to theoretical
predictions.\cite{rolandx}

\noindent{\bf Superlattice Effects}

Next we comment on a possible interplay between a $d$-wave order parameter and
the superlattice modulation in X quadrant. These considerations
were motivated by the fact that observed nodes in
the X quadrant are located at the
same points where the main Fermi surface sheet and
the umklapped side sheets are predicted to cross.
It should be noted at the outset that to the extent the location
of the nodes in the X and Y quadrants are identical (within experimental
resolution), the superlattice cannot be argued to have anything to do with
nodes.
Nevertheless, if one were to focus only on the X quadrant data,
the coincidence noted above would require an explanation.

At this time no firm evidence exists in the data for superlattice effects in
the X
quadrant, primarily because the side sheets are not predicted to be well
separated as they are in the $Y$ quadrant.  Because of this we do not know
yet whether the superlattice modulation which exists on the BiO layers
influences the CuO electronic structure causing a splitting of the
sheets at the crossing points. It is possible that the BiO layer
primarily acts as a diffraction grating for the outgoing
photoelectrons (which would be sufficient to explain the effects
seen in the Y quadrant). Nevertheless, since it will turn out that
only a small superlattice potential on the CuO planes would give the effect
described below, it cannot be apriori excluded.

At the crossing points, then, one has two bands and thus a
$4 \times 4$ secular matrix to diagonalize
(instead of the simple $2 \times 2$ secular matrix in BCS theory).
Schematically this is given by
\begin{eqnarray}
\left(
\begin{array}{cccc}
                \epsilon & V & 0 & \Delta   \\
                V & \epsilon^\prime & s\Delta & 0 \\
                0 & s\Delta & -\epsilon^\prime & -V \\
                 \Delta & 0 & -V & -\epsilon
\end{array}
\right)
\end{eqnarray}
and its eigenvalues
determine the quasiparticle energies in the superconducting state.
Here $\epsilon$ and $\epsilon^\prime$ are the energies of the
two bands measured from the chemical potential,
$V$ is the matrix element of the superlattice potential
mixing the two bands, and
$\Delta$ is the order parameter. The significance of
the parameter $s = \pm 1$ will become clear below.
It is instructive to focus on the analytically solvable
case $\epsilon = \epsilon^\prime$
which corresponds to a locus in ${\bf k}$-space where contours
of constant energy of the two bands intersect.

First consider a gap function  (e.g., an anisotropic s-wave gap)
which is even with respect to reflections about the
$(\pi,\pi)$ direction. This corresponds to $s = +1$, since the
gap is the same for the two sheets at the crossing point.
The eigenvalues (for $\epsilon = \epsilon^\prime$)
are $E = \sqrt{(\epsilon \pm V)^2 + \Delta^2}$.
Thus the superlattice potential just acts to ``shift'' the  normal
state band structure but does not affect the nodes.

For the case of a $d_{x^2-y^2}$ gap, though, the gaps for the two sheets have
{\it opposite} signs at the crossing points, which corresponds to
$s = -1$  In this case, the quasiparticle energies are
$E = \sqrt{\epsilon^2 + \Delta^2} \pm V$.
This is qualitatively different from the previous case
since the superlattice potential can now affect the locations of the nodes:
if $V$ exceeds $\Delta$, a node in $E$ must occur for some value of
$\epsilon$ where there was none previously.

To study the second case in more detail we solve the $6 \times 6$
secular equation describing the main band and the two umklapped side bands
using for each band the appropriate tight binding energy dispersion
in the $X$ quadrant discussed above.  In Fig. 4, we show a plot
of the resulting Fermi surface assuming a superlattice potential of 10 meV
(this potential is assumed to be the same for all off-diagonal terms; we note
that the potential term connecting the two side bands to each other is in
general different from that connecting the main sheet to the side sheets).
In Fig. 5, we plot the resulting excitation gap (minimum
of $E$) on the middle of the three sheets, assuming a
$cos(k_x)-cos(k_y)$ order parameter.
We note the dip in the excitation gap near the observed nodes.  The finite
gap along the $\Gamma-X$ direction occurs since the middle sheet corresponds
to one of the side sheets in this case  (the sheet nearest the $\Gamma$
point along this direction corresponds to the main sheet and has a node; its
gap is also shown in Fig. 5).
Although Fig. 5 has remarkable similarities to Fig. 3, the main problem
for such a model is that it predicts pure $d$-wave behavior for
the excitation gap in the $Y$ quadrant  with a single node along
$\Gamma-Y$, in contrast to experiment.  As noted above, although
the unusual anisotropy of the gap is more obvious in the $X$ quadrant, one
still sees qualitatively similar behavior in the $Y$ quadrant.

\noindent{\bf Pairing within a bilayer}

The final $d$-wave related scenario to consider is the interlayer pairing
model of Ubbens and Lee.\cite{ul}  In this model,
it is assumed that the two CuO bands coming from the bilayer are
degenerate (which is certainly consistent with our data) and
that the pairing within a layer has $d_{x^2-y^2}$ symmetry while
that between the two layers has an anisotropic s-wave symmetry.
Again, one has a $4 \times 4$ secular
matrix to diagonalize for the quasiparticle eigenvalues and the result is two
excitation gaps of the form $d \pm s$.  One of these gaps has a node on
one side of the diagonal direction, the other on the other side.  In Fig. 6,
we show a fit of this theory to the data in the X quadrant assuming the
s-wave component to be isotropic (which is probably a reasonable approximation
on the Fermi surface).  The smaller of the two gaps fits the data
quite well, however, there is no clear signature in the data of the
larger of the two gaps at any point on the Fermi surface.

More specifically, all the spectra can be
fit very well assuming one gap.\cite{gap}
Two gap fits are possible, of course, given
the finite resolution, but our fits to the spectrum at the maximum
gap value (where the quasiparticle peaks are the sharpest) indicates that the
largest splitting between the two gaps that can be accommodated by the
data is about 10 meV.  This should be contrasted with the 18 meV splitting
implied by the fit.
So we would conclude at this stage that the model of Ubbens and Lee is not
inconsistent with the data, but that, so far, our data show no evidence for
a second gap.

\noindent{\bf 5. Anisotropic s-Wave Models}

We now turn to a discussion of anisotropic s-wave pairing, by which
we mean an order parameter which has the full
symmetry of the lattice.
While simple s-wave pairing corresponds to a gap function
independent of ${\bf k}$, i.e., $\Delta({\bf k})=\Delta_0$,
more general s-wave gap functions have the ${\bf k}$-dependence
of the tight binding functions listed in Table 1.
(We will not discuss functions,
such as $\vert \cos(k_x) - \cos(k_y) \vert$, which have
a singular ${\bf k}$-dependence at the node).
Examples of anisotropic s-wave pairing
are the ``extended s-wave'' pairing, denoted by $s^*$
\begin{equation}
\Delta({\bf k}) = \Delta_0 \left( \cos(k_x) + \cos(k_y) \right)
\end{equation}
and $s_{xy}$ pairing
\begin{equation}
\Delta({\bf k}) = \Delta_0 \cos(k_x)\cos(k_y),
\end{equation}
corresponding to the first two nontrivial entries in Table I respectively.
Note that  for the purposes of this discussion we shall treat
Bi-2212 as being tetragonal ignoring the effects of the superlattice.
Fig. 3 is clearly consistent with
this type of order parameter in that the measured gap (a) appears to have
an extremum along the diagonal directions which would require an order
parameter which is even under reflection symmetry about this direction
and (b) has nodes on both sides of the diagonal directions which
is also allowed for this symmetry.

One of the first microscopic examples of anisotropic s-wave pairing was
the $s^*$ solution obtained by Littlewood, Varma, and Abrahams,\cite{varma}
in the context of charge fluctuation
mediated pairing in a three-band model.
Depending on the precise shape of the Fermi surface an $s^*$ gap
has either no nodes (i.e., fully gapped) or two nodes per quadrant.\cite{fedro}
A more general phenomenological analysis was presented by Chen and
Tremblay \cite{tremblay} who also discussed, among other things, the nodal
structure of the $s_{xy}$ state which will be shown to be relevant to the
ARPES data.

Before discussing microscopic models, we first
describe phenomenological fits to the ARPES gap
data assuming a sign change in the gap function
at the node observed in the experiment (so that the
gap function is smooth at the node).  In Fig. 7, we plot the location of the
nodes of the $s^*$ and $s_{xy}$ gap in the Brillouin zone.  We note that
the nodes of the $s_{xy}$ gap function are very close to the experimentally
observed nodes in the excitation spectrum,
whereas those of the $s^*$ gap function are
significantly further from the diagonal direction compared with the data.
This can be seen more clearly in Fig. 8 where we plot these two functions
over the observed Fermi surface.
In fact, we find, quite surprisingly, that a pure $s_{xy}$ gap
fits the data very well in the $X$ quadrant as shown in Fig. 9.
A least squares fit of the form $s^* + s_{xy}$ gives only a
$2\%$ admixture of $s^*$ and one of the form $s + s_{xy}$ gives only a $0.2\%$
admixture of the constant ($s$) term.
Alternatively an $s + s^*$ gap can also fit the location of the
nodes, but at the expense of having too small a gap along the diagonal
directions (this fit has the two terms comparable in size but with
opposite sign).

Within a simple, single band BCS framework, given
an effective electron-electron interaction $ - V_{{\bf k},{\bf k}^\prime}$,
the gap function at a temperature $T$ is determined by solving the
self-consistent equation
\begin{equation}
\Delta({\bf k}) = \sum_{{\bf k}^\prime} V_{{\bf k},{\bf k}^\prime}
{\Delta({\bf k}^\prime) \over 2E({\bf k}^\prime)}
\tanh\left[{E({\bf k}^\prime)\over 2 T}\right],
\end{equation}
where $E({\bf k}) = \sqrt{\epsilon^2({\bf k}) + \Delta^2({\bf k})}$.
Here we will ``invert'' this procedure and infer
some features of $V_{{\bf k},{\bf k}^\prime}$ which would be consistent
with the observed $\Delta({\bf k})$. The simplest potentials which
give rise to $s$, $s^*$, and $s_{xy}$ pairing
are attractive on-site, on the near neighbor (NN) site,
and on the next near neighbor (NNN) site respectively.

The fact that a constant ($s$) term is not needed in the fit is significant in
that it suggests that the large Coulomb repulsion on the Cu site is
substantially renormalized in the pairing interaction between quasiparticles.
We note that although the gap functions discussed above, other than the simple
$s$ state, satisfy the condition $\sum_{\bf k} \Delta({\bf k}) = 0$, it is not
this condition which must be satisfied to eliminate the effect of an on-site
repulsion in the gap equations.
Rather, it is the condition $\sum_{\bf k}\Delta({\bf k})/2E({\bf k})
\tanh\left[E({\bf k})/2T\right] = 0$ which needs to be satisfied.  Clearly, a
pure $s_{xy}$ function violates this condition which implies such
a repulsive term in the gap equations must be small.\cite{roland}

The dominance of the NNN attraction in the singlet channel
suggested by the $s_{xy}$ pairing is also very interesting.
To avoid any confusion, we emphasize that NNN means the NNN Cu site,
and not the pairing of holes on neighboring oxygen sites.
(The $s_{xy}$ order parameter $\cos(k_x)\cos(k_y)$ Fourier transformed
to real space has peaks in the
relative coordinate near the $(\pm 1,\pm 1)$ primitive lattice vectors).
In most theories proposed to date,
NNN pairing terms are weak and would play a
minor role; in fact, we know of no theory where such
terms determine the symmetry of the gap.

We now turn to microscopic models. The charge transfer model of Varma
and coworkers with $s^*$ pairing was already mentioned earlier.
The valence fluctuation model of Brandow\cite{baird} also gives rise to
anisotropic s-wave pairing where the existence and location of nodes
depends on the details of the input parameters and on the Fermi surface.

We finally discuss two other theories: the interlayer tunneling
model of Chakravarty et al.\cite{sudip} and the recent theory of
Abrikosov\cite{alex,alex2} based on a weakly screened
electron-phonon interaction in the presence of extended saddle point
singularities.
While the physics underlying the two theories is completely different
they share two characteristics: first, the dominant part of the
interactions are ``local in ${\bf k}$-space'' (for entirely different reasons),
and second, both theories have the ability of having different order parameter
symmetries in different materials with comparable $T_c$
unlike in simple BCS models.

In Abrikosov's theory the poor screening of
the electron-phonon interaction causes a sharp peak in the forward
scattering direction (small momentum transfers).
Further the local density of states is very large near
the extended saddle point singularity  or flat band,
around the $\bar{M}$ point in Bi-2212, and small away
from it. The solution of the gap equation thus yields
a highly anisotropic gap function which is large near
$\bar{M}$ point and small away from it.
An additional short range
Coulomb repulsion is required to produce nodes
as pointed out recently by Abrikosov.\cite{alex2}
This adds an overall negative constant to the order parameter,
thus causing the small order parameter along
the diagonal direction to become negative, with nodes on
two sides of the diagonal in a tetragonal material like Bi-2212.
However in an orthorhombic material like YBCO
which in this theory is modeled as having an extended
saddle point only along one direction,
Abrikosov finds an $s+d$ gap with a sign change going
from the $a$ to the $b$ axis.

The Chakravarty et al. model is based on ideas of Anderson\cite{pwa} that
the elementary excitations, the holons and spinons,
are confined within a plane, and thus the coherent transport of single
electrons from one plane to another is not possible (this is certainly
consistent with our observation of only one CuO band).
The dominant interaction
is the tunneling of Cooper pairs between layers
which lowers the kinetic energy dramatically and thus acts to drive
the superconducting transition.
This pair tunneling term, which is assumed to be
completely local in momentum space, leads to the
high $T_c$ but does not constrain the symmetry of the order parameter.
The in-plane BCS interaction, which is a weak additive term,
is as usual non-local in momentum variables and thus
determines the symmetry of the gap, with the momentum dependence of this gap
being modulated by the ${\bf k}$-dependence of the tunneling
term $t_\perp({\bf k})$. (In Bi-2212  electronic structure
calculations suggest that  $t_\perp({\bf k})$ is large along the CuO
bond direction and small along the diagonal direction).
In the Chakravarty et al. model, the largest of the weak in-plane BCS terms
could differ from one material to another, thus leading
to the possibility mentioned in
the beginning of this paper of different pairing symmetries
for different cuprates.

{\bf 6. Conclusions}

In conclusion, we have analyzed various theoretical models in light of
our measured gap anisotropy in Bi-2212.  We find that a simple d-wave model
can be made consistent with the data in the $X$ quadrant, but not in the
$Y$ quadrant, if the superlattice modulation is taken into account.
We find that the only d-wave model consistent with the observed behavior
in both quadrants is an interlayer pairing model \cite{ul}
although we remark that so far, we have no evidence for
the two-gap spectrum required by such a model.
The simplest gap function which naturally explains the two nodes
observed per quadrant is an $s_{xy}$ order parameter of the form
$\Delta_0 \cos(k_x)\cos(k_y)$. Within a one band BCS framework such
an order parameter is obtained from a next-near neighbor (Cu-Cu)
interaction. Several microscopic theories \cite{varma,baird,sudip,alex,alex2}
which give rise to anisotropic s-wave pairing were discussed in this context.

\acknowledgements

We would like to thank Alex Abrikosov, Phil Anderson, Baird
Brandow, Sudip Chakravarty, Roland Fehrenbacher, Patrick Lee,
and Chandra Varma for helpful discussions.
This work was supported by the U.~S.~Department of Energy,
Basic Energy Sciences, under Contract \#W-31-109-ENG-38.

\begin{table}
\caption{Tight binding basis functions used in
fitting the experimental energy dispersions as described in the text.
In this notation, the $Y$ point is $(\pi,\pi)$.
The first column lists the coefficient of each term (eV), that is
$\epsilon(\vec k) = \sum c_i \eta_i(\vec k)$.}
\begin{tabular}{cc}
$c_i$ & $\eta_i(\vec k)$ \\
\tableline
 0.1305 & $1$ \\
-0.5951 & $\frac{1}{2} (\cos k_x + \cos k_y)$ \\
 0.1636 & $\cos k_x \cos k_y $ \\
-0.0519 & $\frac{1}{2} (\cos 2 k_x + \cos 2 k_y)$ \\
-0.1117 & $\frac{1}{2} (\cos 2k_x \cos k_y + \cos k_x \cos 2k_y)$ \\
 0.0510 & $\cos 2k_x \cos 2k_y $ \\
\end{tabular}
\end{table}

\begin{figure}
\caption{Fermi surface points measured by ARPES.  Open circles were
measurements in the normal state, crosses in the superconducting
state.  The thick line represents a tight binding fit to the data as
described in the text.  The thin lines are displaced from this by the
superlattice Q vector.}
\label{fig1}
\end{figure}

\begin{figure}
\caption{Energy dispersion from the tight binding fit.  The filled circles
were data points used in the fit (the Y point being obtained as described
in the text).  Note the
extended van Hove singularity at the $\bar{M}$ point.}
\label{fig2}
\end{figure}

\begin{figure}
\caption{Measured energy gap along the Fermi surface in the $X$ and $Y$
quadrants.  The data for each quadrant were taken on a different
sample.  In this notation, 0 degrees corresponds to the $\bar{M}-X,Y$
directions and 45 degrees to the $\Gamma-X,Y$ directions, with the angle
measured with respect to an origin at the $X,Y$ points.}
\label{fig3}
\end{figure}

\begin{figure}
\caption{Fermi surface determined by the tight binding fit with an
additional superlattice potential of 10 meV.  Data points as in Fig. 1.}
\label{fig4}
\end{figure}

\begin{figure}
\caption{Calculated excitation gap in the $X$ quadrant assuming a $d_{x^2-y^2}$
order parameter with a coefficient of 12.5 meV and a superlattice potential of
10 meV on the middle of the three Fermi surface sheets shown in Fig. 4
(filled circles) and on
the inner sheet nearest $\Gamma$ (open circles).}
\label{fig5}
\end{figure}

\begin{figure}
\caption{Fit of the experimental data in the $X$ quadrant to the smaller of
the two gaps of the Ubbens and Lee model.  Filled circles are the data and
pluses (+) the two gaps of this model.}
\label{fig6}
\end{figure}

\begin{figure}
\caption{Plots of the nodes of the $s^*$ and $s_{xy}$ functions in the
Brillouin zone (dashed lines).  The former run diagonally, the latter
horizontally and vertically.  The solid line marks the Fermi surface.}
\label{fig7}
\end{figure}

\begin{figure}
\caption{Plots of the $s^*$ and $s_{xy}$
anisotropic s-wave functions
along the Fermi surface in the $X$ quadrant of the zone.}
\label{fig8}
\end{figure}

\begin{figure}
\caption{Fit of the $s_{xy}$ anisotropic s-wave gap to
the data in the $X$ quadrant.  Filled circles are the data and pluses (+)
the fit.}
\label{fig9}
\end{figure}


\begin{references}

\bibitem{hardy} W. N. Hardy et al., Phys. Rev. Lett. {\bf 70}, 3999 (1993).

\bibitem{vh} D. A. Wollman et al., Phys. Rev. Lett. {\bf 71}, 2134 (1993)
and preprint;
D. A. Brawner and H. R. Ott, Phys. Rev. B {\bf 50}, 6530 (1994);
A. Mathai et al., preprint.

\bibitem{ring} C. C. Tsuei et al., Phys. Rev. Lett. {\bf 73}, 539 (1994);
J. R. Kirtley et al., preprint.

\bibitem{dynes} A. G. Sun et al., Phys. Rev. Lett. {\bf 72}, 2267 (1994).

\bibitem{chaud} P. Chaudhari and S.-Y. Lin, Phys. Rev. Lett. {\bf 72},
1084 (1994) and preprint.

\bibitem{anlage} D. H. Wu et al., Phys. Rev. Lett. {\bf 70}, 85 (1993).

\bibitem{jz} Q. Huang et al., Nature {\bf 347}, 369 (1990).

\bibitem{jz2} J. Chen et al., Phys. Rev. B {\bf 49}, 3683 (1994).

\bibitem{shen} Z.-X. Shen et al., Phys. Rev. Lett. {\bf 70}, 1553 (1993).

\bibitem{ding} H. Ding et al., Phys. Rev. B {\bf 50}, 1333 (1994).

\bibitem{onel} R. J. Kelley et al., Phys. Rev. B {\bf 50}, 590 (1994).

\bibitem{olson} C. G. Olson et al., Solid State Comm. {\bf 76}, 411 (1990).

\bibitem{bansil} A. Bansil et al., J. Phys. Chem. Solids {\bf 54}, 1185 (1993).

\bibitem{sample} T. Mochiku and K. Kadowaki, in Proceedings of the
IUMRS-ICAM-93, Tokyo, 1993, to be published.

\bibitem{gap} H. Ding et al., Momentum Dependence of the superconducting gap
in BSCCO, preprint.

\bibitem{band} S. Massidda, J. Yu, and A. J. Freeman, Physica C {\bf 152},
251 (1988); H. Krakauer and W. E. Pickett, Phys. Rev. Lett. {\bf 60},
1665 (1988).

\bibitem{olson2} C. G. Olson et al., Phys. Rev. B {\bf 42}, 381 (1990).

\bibitem{shen2} D. S. Dessau et al., Phys. Rev. Lett. {\bf 71}, 2781 (1993).

\bibitem{pwa} P. W. Anderson, Science {\bf 256}, 1526 (1992) and
``A Career in Theoretical Physics'', p.~638, (World Scientific, 1994).

\bibitem{kaz} K. Gofron et al., J. Phys. Chem. Solids {\bf 54}, 1193 (1993)
and Phys. Rev. Lett. {\bf 73}, xxxx (1994).

\bibitem{oka} O. K. Andersen, O. Jepsen, A. I. Liechtenstein, and I. I.
Mazin, Phys. Rev. B {\bf 49}, 4145 (1994).

\bibitem{with} R. L. Withers et al., J. Phys. C {\bf 21}, 6067 (1988).

\bibitem{aebi} P. Aebi et al., Phys. Rev. Lett {\bf 72}, 2757 (1994).

\bibitem{scalapino} N. E. Bickers, D. J. Scalapino, and R. T. Scaletar,
Intl. J. Mod. Phys. B {\bf 1}, 687 (1987); T. Moriya, Y. Takahashi, and K.
Ueda, J. Phys. Soc. Japan {\bf 59}, 2905 (1990); P. Monthoux, A. V. Balatsky,
and D. Pines, Phys. Rev. Lett. {\bf 67}, 3448 (1991).

\bibitem{takigawa} M. Takigawa and D. B. Mitzi,
Phys. Rev. Lett {\bf 73}, 1287 (1994).

\bibitem{kwj} P. Kumar and P. Wolfle, Phys. Rev. Lett. {\bf 59}, 1954 (1987);
Q. P. Li, B. E. C. Koltenbah, and R. Joynt, Phys. Rev. B {\bf 48}, 437 (1993).

\bibitem{rolandx} R. Fehrenbacher and M. R. Norman, Phys. Rev. B {\bf 50},
3495 (1994).

\bibitem{ul} M. U. Ubbens and P. A. Lee, Phys. Rev. B {\bf 50}, 438 (1994).

\bibitem{varma} P. B. Littlewood, C. M. Varma, and E. Abrahams, Phys. Rev.
Lett. {\bf 63}, 2602 (1989); P. B. Littlewood, Phys. Rev. B {\bf 42}, 10075
(1990).

\bibitem{fedro} A. J. Fedro and D. D. Koelling,
Phys. Rev. B {\bf 47}, 14342 (1993).

\bibitem{tremblay} L. Chen and A.-M. S. Tremblay,
J. Phys. Chem. Solids {\bf 54}, 1381 (1993).

\bibitem{roland} R. Fehrenbacher and M. R. Norman, preprint.

\bibitem{baird} B. H. Brandow, Intl. J. Mod Phys. B {\bf 8}, 2667 (1994).

\bibitem{sudip} S. Chakravarty, A. Subdo, P. W. Anderson, and S. Strong,
Science {\bf 261}, 337 (1993).

\bibitem{alex}A. A. Abrikosov, Physica C {\bf 222}, 191 (1994).

\bibitem{alex2}A. A. Abrikosov, preprint.

\end{references}
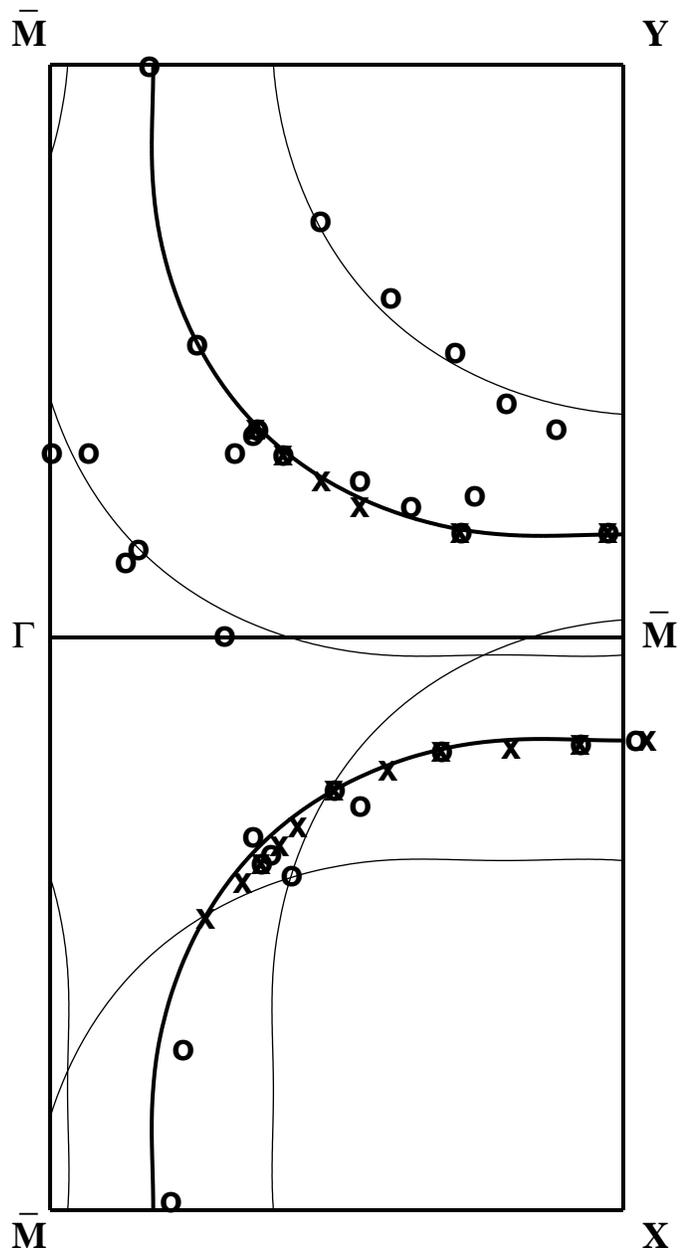
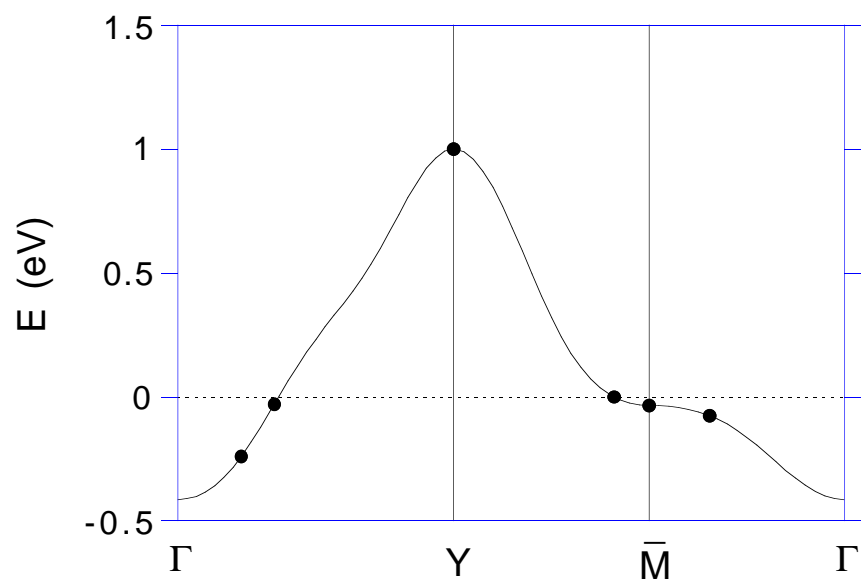
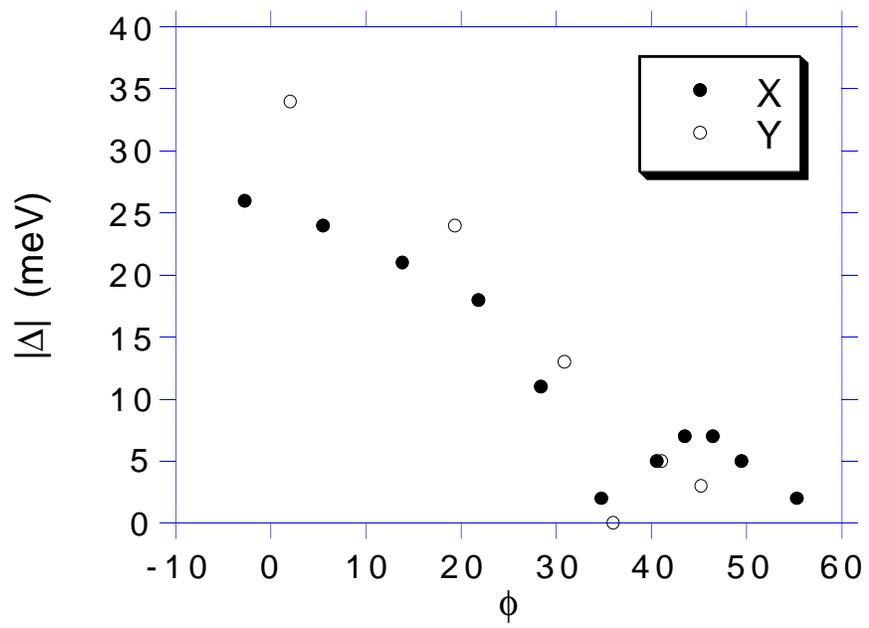
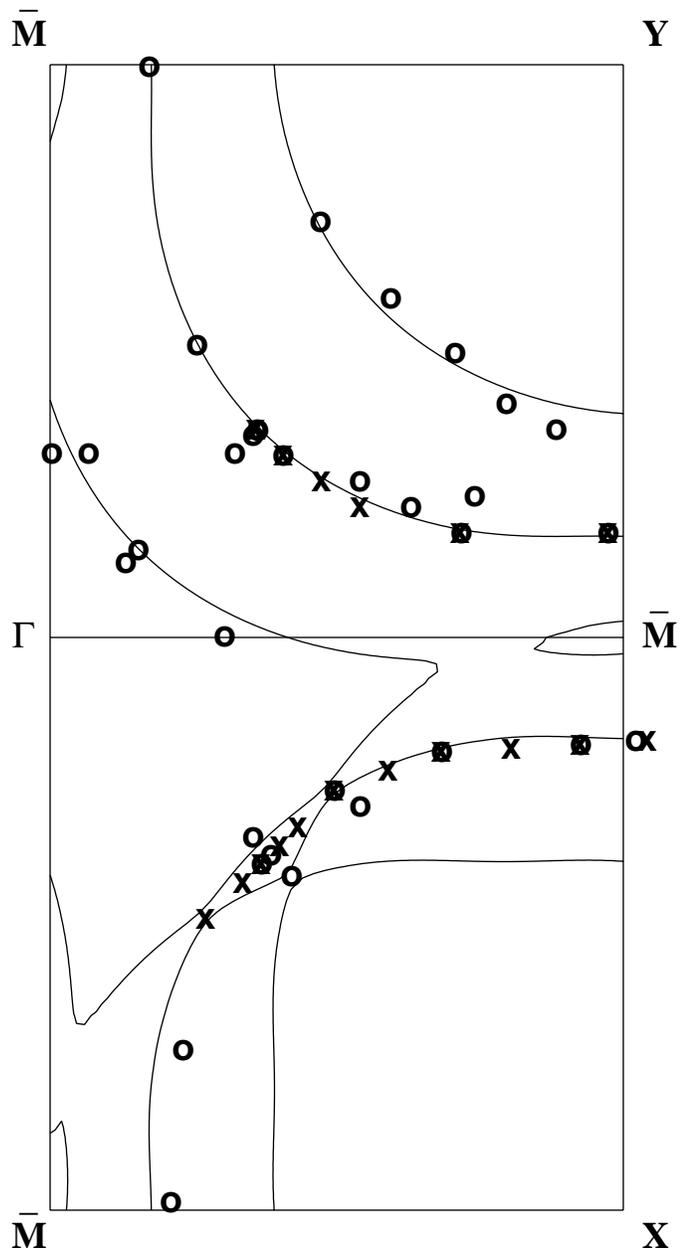
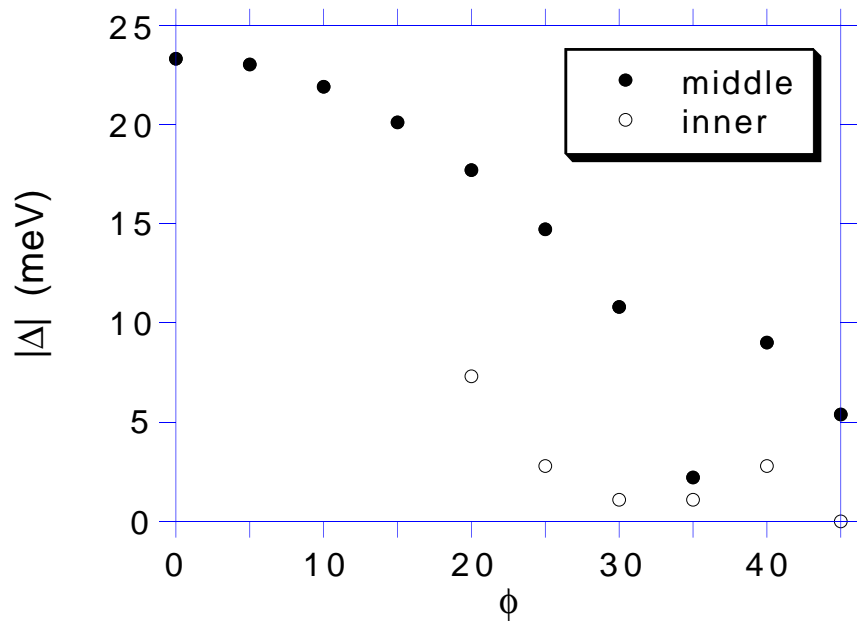
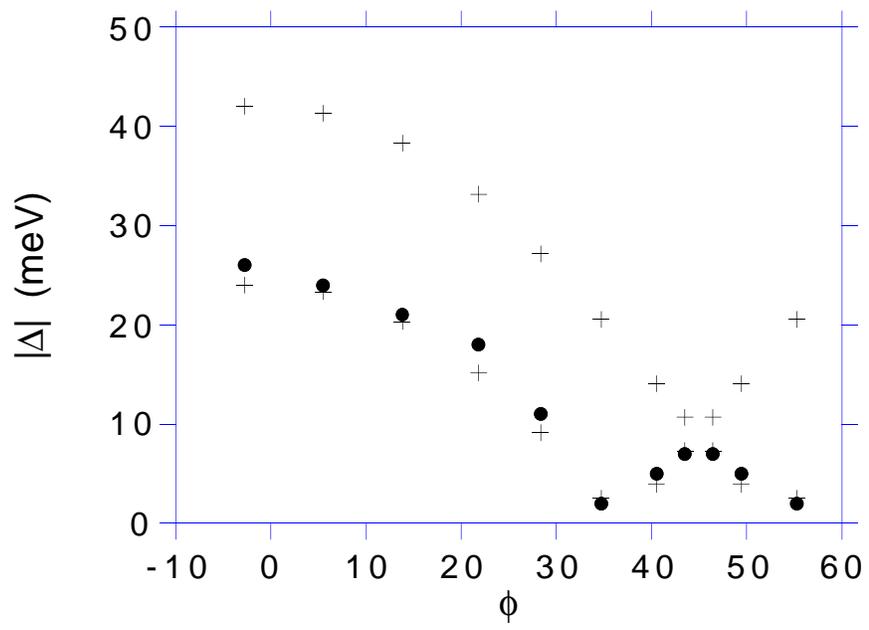
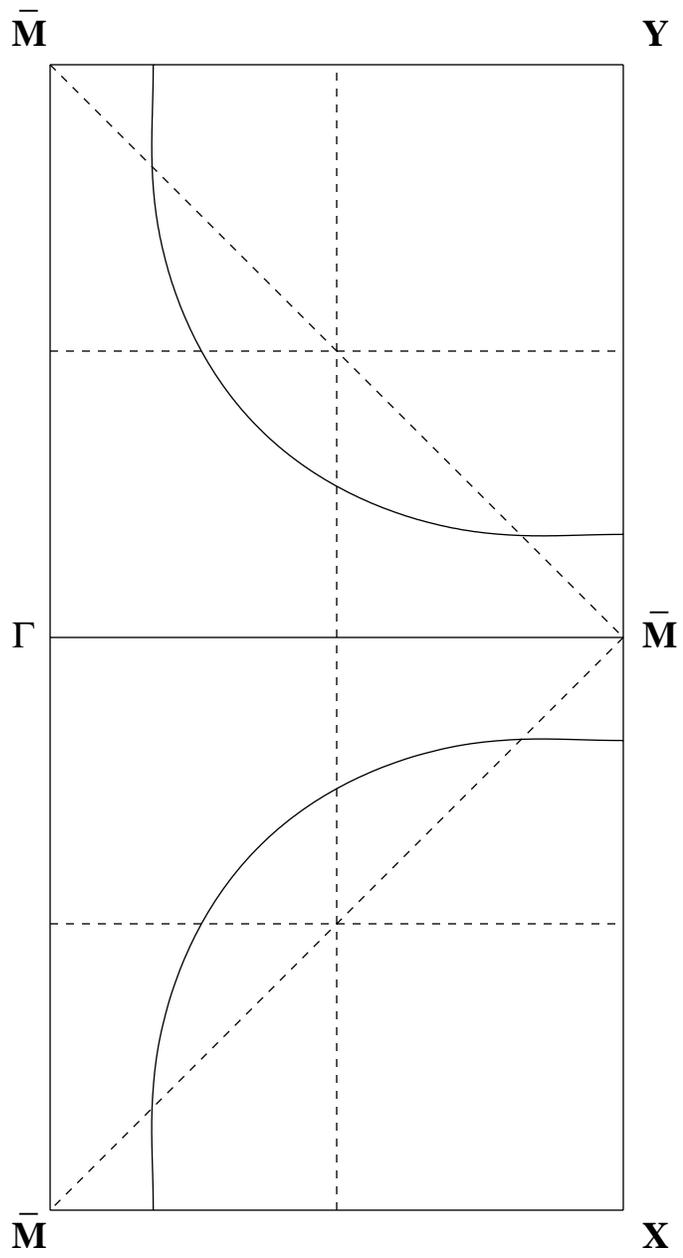
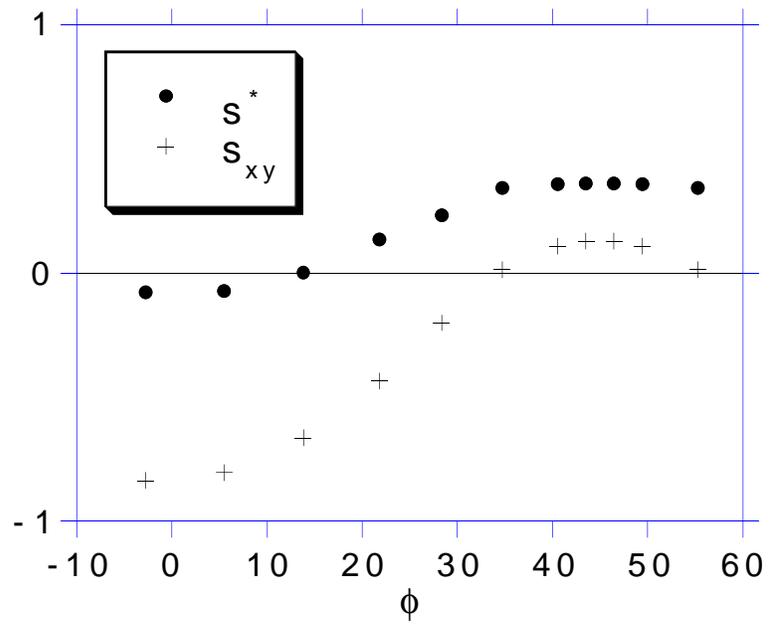
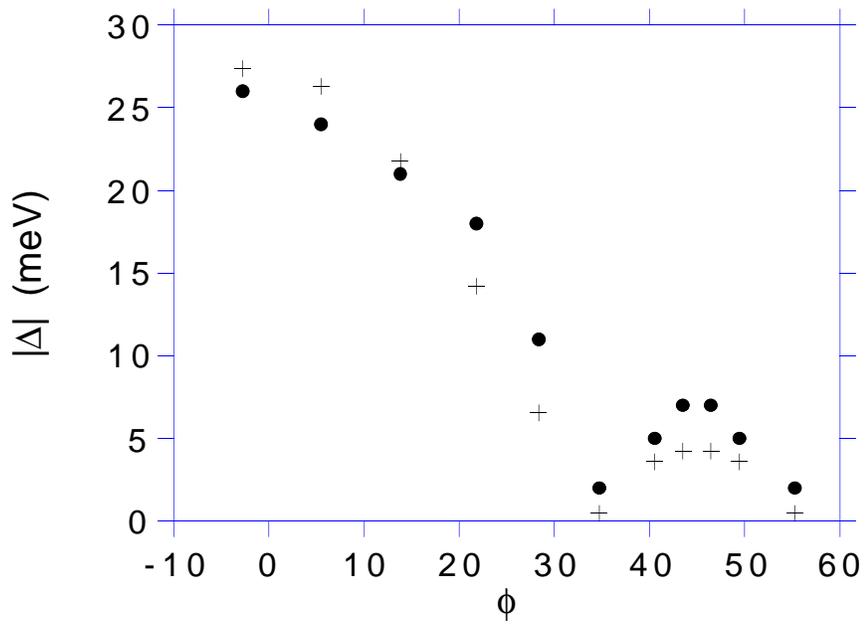
\end{document}